\documentclass [11pt] {article}
\usepackage[papersize={8.5in,11in},hdivide={1in,*,1in},vdivide={1in,*,1in},dvips,pdftex]{geometry}
\usepackage{indentfirst,latexsym,bm}
\usepackage{setspace}
\usepackage{amsmath}
\usepackage{amssymb}
\usepackage{hyperref}
\usepackage{amsthm}
\usepackage{epsfig}
\usepackage{graphicx}
\usepackage{graphics}
\usepackage[normal]{subfigure}
\usepackage{amsfonts}

\usepackage[referable]{threeparttablex}

\usepackage{float}

\title{A Sketch of Some Stochastic Models and Analysis Methods for Fiber Bundle Failure under Increasing Tensile Load}
\author{
        Shuang Li\footnote{Department of Mathematics, Florida Gulf Coast University, 10501 FGCU Boulevard South, Fort Myers, FL 33965-6565, sli@fgcu.edu} \mbox{ } and James Lynch\footnote{Department of Statistics, University of South Carolina, Columbia, SC 29208, USA. lynch@stat.sc.edu}
        }

\date{\today}

\begin{document}
\maketitle
\begin{abstract}
Fiber bundle models (FBM's) have been used to model the failure of fibrous composites as load-sharing systems since the 1960's when Rosen (1964 and 1965) conducted some remarkable experiments on unidirectional fibrous composites.  These experiments gave seminal insights into their failure under increasing tensile load.  However, over the last thirty years FBM's have been used to model catastrophic failure in other situations by the physical science community and others.  The purpose of this paper is to sketch some research on load-sharing models and statistical analysis methods that have been overlooked by this community.  These are illustrated by summarizing the findings regarding Rosen's Specimen A experiments and presenting the necessary results needed for this. Related research about the bundle breaking strength distribution and the joint distribution (the Gibbs measure) regarding the state (failed or unfailed) of the bundle components at a given load per component, $s$, is also given.

\noindent \begin{bfseries}\begin{itshape}keywords:\end{itshape}\end{bfseries} {Fiber Bundle Model, Chain-of-Bundles, Load-Sharing Rule, Bundle Strength Distribution, Gibbs Measure, Potentials, Weibull Analysis, Minimum Laws}

\end{abstract}
\newpage

\section{Introduction}
Fiber bundle models (FBM's) have been used to model the failure of fibrous composites as load-sharing systems since the 1960's when Rosen (1964 and 1965) conducted some remarkable experiments on unidirectional fibrous composites.  These experiments gave seminal insights into their failure under increasing tensile load.  However, over the last thirty years FBM's have been used to model catastrophic failure in other situations (Kun et al., 2006, and Hansen et al., 2015) by the physical science community and others.

Kun et al. (2006) is one of several introductory chapters in a book cited in the references where the chapters after the introduction are specialized applications of the FBM in the geosciences.   On the other hand Hansen et al. (2015)'s book is an excellent introduction to FBM's that is accessible to nonphysicists.  The purpose of this sketch is not to give a comprehensive review of the literature (the latter two references and Li et al., 2018, are good sources to start this) but rather to sketch some research on load-sharing models and statistical analysis methods relevant for FBM's that have been overlooked by the  physical science community.  These are illustrated by summarizing the findings from Li et al. (2018) regarding Rosen's Specimen A experiments and the necessary results needed for this.  Related research about the bundle breaking strength distribution and the joint distribution (the Gibbs measure) regarding the state (failed or unfailed) of the bundle components at a load per component, $s$, is also given.

	In Section 2, a threshold distribution (a scale mixture) is given for the bundle breaking strength when the bundle component strengths are independent Weilbulls.  The bundle strength distribution is a gamma type mixture over the gamma scale parameter where the scale mixing distribution is totally determined by the load-sharing rule.  In Section 3, the construction of the joint distribution of the components' states is given.  It is a Gibbs measure that has a simple local structure given in (3.6) in terms of the log odds of a component strength distribution evaluated based on the load-sharing rule.

	Rosen's Series A specimen experiments are described in Section 4.  The nonparametric Bayes estimation of the component strength distribution is given in Section 4.2 where a partition based prior is used because the component breaking strengths from Rosen's photographs of Specimen A-7 are all censored.  The local structure of the load-sharing that was used in the analysis of the photographs sufficed for this purpose but did not define a load-sharing rule.  This is needed to model the Series A Specimen breaking strengths as a chain-of-grid-bundles in Section 4.5 and to define the Gibbs measure for the $4\times4$ grid-bundle in Section 4.6.  In Section 4.3, it is shown that, when the local structure of the load-sharing defines one step transition probabilities for a Markov chain, absorbing state probabilities for the chain can be used to define a monotone load-sharing rule based on this local structure.
	
	To keep the paper self-contained, a simple presentation of the extreme value asymptotics needed in Section 4.5 is given in Section 4.4. (A brief but complete theory for minimum extreme value asymptotics can be found in Chapter 8 of Barlow and Proschan, 1975.)  With regard to chain-of-bundles asymptotics, Harlow et al. (1983) addressed this for bundles that were parallel systems of $k$ components under monotone load-sharing.  They showed that, when the component strength distribution behavior was Weibull-like with shape parameter $\rho$ at the origin, the chain breaking strength was asymptotically Weibull with shape parameter $k\rho$ where $k$ is referred to, later, as the \begin{bfseries}\begin{itshape}inflation factor\end{itshape}\end{bfseries}. They also noted that much of the load-sharing used in the literature was local in structure and did not lend itself for asymptotic purposes; you needed to have a load-sharing rule that prescribes the load-sharing for all possible configurations of unfailed components.  Their asymptotic results were derived for load-sharing rules that were what they called \begin{bfseries}\begin{itshape}monotone load-sharing rules\end{itshape}\end{bfseries}. Lynch (2001) considered the same situation, but for Weibull components and, as such, was able to exploit the bundle threshold representation given in Durham and Lynch (2000).  Using this representation, he was able to obtain not just limit results but approximations and investigated not just parallel systems but arbitrary systems.  In this case the inflation factor is the size of the smallest cut set defined later.

Finally, in Section 5, we have two concluding comments regarding FBM's.  The first is in regards to the Markovian behavior of the component failures under increasing load.  The second describes how, with certain assumptions on component degradation, the results in the previous sections might be pieced together to develop a theory and methods to analyze cycles to failure testing data.

The following notation is used in later sections.

\noindent \begin{bfseries}Notation:\end{bfseries}\\
\begin{bfseries}(i)\end{bfseries}  $X\sim f$ or $X\sim F$ denotes that the random variable $X$ has distribution $F$ or density $f$.\\
\begin{bfseries}(ii)\end{bfseries} In a different context than (i), $A_n \sim B_n$ means $A_n / B_n \overset{n\rightarrow \infty}{\longrightarrow} 1$.\\
\begin{bfseries}(iii)\end{bfseries} $A(\varepsilon)=o(\varepsilon)$ means $A(\varepsilon)/\varepsilon \overset{\varepsilon \rightarrow 0}{\longrightarrow} 0$.\\

\section {Threshold Representations}

In Section 2.1 we give threshold representations for the order statistics of iid exponential random variables.  It is well known that the asymptotic distribution of the $k^{th}$ order statistic is a gamma with shape parameter $k$.  A consequence of the representations given here is that, for a random sample of size the $n$, the $k^{th}$ order statistic's exact distribution is a scale mixture of gammas with shape parameter $k$ where the mixing is over the scale parameter and the mixing distribution is based on a convolution of $k-1$ uniforms and the sample size $n$.

	Similar representations are given in Section 2.2 for the breaking strengths in a fiber bundle.  Here the mixing distributions are also convolutions of uniforms which are based on the load-sharing rule for bundle and the shape parameter is related to the number of fiber components that break in the bundle.

\subsection {A Threshold Representation for Order Statistics}

Following Grego and Lynch (2006), let $X_1, X_2, \ldots, X_n$ be iid random variables with distribution $F$ and density $f$. Denote their ordered values by $X_{1;n} < X_{2;n} < \cdots < X_{n;n}$. Then, for $1\leq k < l \leq n$ and $0 < x < y$, $(X_{k;n}, X_{l;n}) \sim f_{k,l,n}(x,y)$, where

$$f_{k,l,n}(x,y)\propto (F(x))^{k-1}f(x)(F(y)-F(x))^{l-k-1}f(y)(\bar{F}(y))^{n-l}. \eqno(2.1.1a)$$

When $X_1, X_2, \ldots, X_n$ and iid unit exponentials (i.e., $\bar{F}(x)\equiv 1-F(x) = e^{-x})$,
\begin{align}
 f_{k,l,n}(x,y) &\propto (1-e^{-x})^{k-1}e^{-x}(e^{-x}-e^{-y})^{l-k-1}e^{-y}(e^{-y})^{n-l}\nonumber\\
                &= (1-e^{-x})^{k-1}e^{-(l-k)x}(1-e^{-(y-x)})^{l-k-1}e^{-(n-l+1)y}\tag{2.1.1b}\\
                &= (1-e^{-x})^{k-1} e^{-(n-k+l)x}(1-e^{-(y-x)})^{l-k-1}e^{-(n-l+1)(y-x)} \nonumber
\end{align}

Since $$\frac{1-e^{-x}}{x}=\int_{0}^{1} e^{-\theta x} d\theta , \eqno(2.1.1c)$$
$$\frac{(1-e^{-x})^{k-1}}{x^{k-1}}=\int_{0}^{k-1} e^{-\theta x}b_{k-1}(\theta) d\theta \eqno(2.1.1d)$$
and
$$\frac{(1-e^{-(y-x)})^{l-k-1}}{(y-x)^{l-k-1}}=\int_{0}^{l-k-1} e^{-\theta (y-x)}b_{l-k-1}(\theta) d\theta\eqno(2.1.1e)$$
where $b_{m}(\theta)$ denotes the convolutions of $m$ uniforms on $[0,1]$. So,
$$f_{k,l,n}(x,y)=\iint \frac{\theta_{1}^{k} x^{k-1} e^{-\theta_{1}x}}{(k-1)!} \frac{\theta_{2}^{l-k} (y-x)^{l-k-1} e^{-\theta_{2}(y-x)}}{(l-k-1)!}a_{k;n}(\theta_{1})a_{l-k;n}(\theta_{2})d\theta_{1}d\theta_{2} \eqno(2.1.2a)$$
where $$ a_{k;n}(\theta) \propto \frac{b_{k-1}(\theta-(n-k+1))}{\theta^{k}} \mbox{ and } a_{l-k;n}(\theta) \propto \frac{b_{l-k+1}(\theta-(n-l+1))}{\theta^{l-k}}. \eqno(2.1.2b)$$
Thus, $a_{k;n}(\theta)$ and  $a_{l-k;n}(\theta)$ are probability densities that are positive on the intervals $[n-k+1,n]$ and $[n-l+1,n]$, respectively, and vanish off their complements; they are size biased, shifted convolutions of uniforms. They also quantify the error in the gamma approximation to $f_{k,l;n}(x,y)$.

The mixture representation given in (2.1.2) for order statistics is related to representations for load-sharing systems given in Lynch (1999) and Durham and Lynch (2000) where the mixing distribution was not specified as it was done here for order statistics.  The complete specification for load-sharing systems was given, though, in Li and Lynch (2011).

Let $X, Y, \Theta_{1}$ and $\Theta_{2}$ denote the random variables given by the construction in (2.1.2).  Then, from (2.1.2a), the joint density of these random variables is

$$f_{X, Y, \Theta_{1}, \Theta_{2}}(x,y, \theta_{1}, \theta_{2})= \frac{\theta_{1}^{k} x^{k-1} e^{-\theta_{1}x}}{(k-1)!} \frac{\theta_{2}^{l-k} (y-x)^{l-k-1} e^{-\theta_{2}(y-x)}}{(l-k-1)!}a_{k;n}(\theta_{1})a_{l-k;n}(\theta_{2}) \eqno(2.1.3)$$
where $(X,Y) \sim f_{k,l;n}(x,y)$. It follows from (2.1.3) that $\Theta_{1} \sim a_{k;n}(\theta_{1})$ and $\Theta_{2} \sim a_{l-k;n}(\theta_{2})$.

It also follows from (2.1.3) that $X$ and $Y$ given $\Theta_{1}$ and  $\Theta_{2}$ have joint conditional density

\begin{align}
f_{X, Y \mid \Theta_{1}, \Theta_{2}}(x,y \mid \theta_{1}, \theta_{2}) &= \frac{\theta_{1}^{k} x^{k-1} e^{-\theta_{1}x}}{(k-1)!} \frac{\theta_{2}^{l-k} (y-x)^{l-k-1} e^{-\theta_{2}(y-x)}}{(l-k-1)!}\nonumber\\
                &= f_{X \mid \Theta_{1}}(x \mid \theta_{1}) f_{Y \mid X, \Theta_{2}}(y \mid x, \theta_{2})\tag{2.1.4}
\end{align}

From (2.1.2b) and (2.1.3), the joint conditional density of $\Theta_{1}$ and  $\Theta_{2}$  given $X$ and $Y$  is
\begin{align}
f_{\Theta_{1}, \Theta_{2}\mid X, Y}(\theta_{1}, \theta_{2}\mid x,y) &= \frac{e^{-x\theta_{1}}b_{k-1}(\theta_{1}-(n-k+1))}{l_{k-1}(x)} \frac{e^{-(y-x)\theta_{2}}b_{l-k+1}(\theta_{2}-(n-l+1))}{l_{l-k-1}(y-x)}\nonumber\\
&= f_{\Theta_{1} \mid X}(\theta_{1} \mid x) f_{\Theta_{2} \mid X,Y }( \theta_{2} \mid x,y)\tag{2.1.5}
\end{align}
where the normalizing constants in the denominator are in terms of the Laplace transforms of the convolutions of uniforms.  The term in the middle identity of (2.1.5) is a product of these exponentially tilted convolutions and shows that $\Theta_{1}$ and  $\Theta_{2}$  are conditionally independent given $X$ and $Y$.

In the next section similar distributional results to those given in (2.1.2-5) are excerpted from Li and Lynch (2011), Section 3 for load-sharing systems (bundles).  See that reference for all the details.

\subsection {Threshold Representations Related to the Bundle Strength}

In this section, we derive the threshold representations of the joint distribution of the Phase I failure strengths and of the system strength for a monotone load-sharing system and some consequences of these representations.

The strength of a load-sharing system is based on the \begin{bfseries}\begin{itshape}nominal load per component\end{itshape}\end{bfseries} (the load per component), say $s$. We assume that there are $n$ components in a system which are labeled by $N = \{1, 2, \ldots, n\}$.  Let $M \subseteq N$ denote the set of working components in $N$. Then the load at component $i \in M$ for a nominal load per component $s$ is given by $\lambda_{i}(M)s$. These non-negative constants, $\lambda_{i}(M)$, quantify the load-sharing and the collection $\{\lambda_{i}(M): i\in M, M\subseteq N\}$ is called a load-sharing rule.

When a system of components is subjected to an increasing load, the functioning components share the load according to the load-sharing rule for the system. As the load is increased, a component fails if its strength is less than the load given by the load-sharing rule. There are two types of failures for an individual component. When a component fails directly due to the increasing load, it is referred as a \begin{bfseries}\begin{itshape}Phase I failure\end{itshape}\end{bfseries}. We refer to the nominal load that causes a Phase I failure as a \begin{bfseries}\begin{itshape}Phase I breaking stress\end{itshape}\end{bfseries}.  A Phase I failure can initiate a series of additional instantaneous failures (referred as \begin{bfseries}\begin{itshape}Phase II failures\end{itshape}\end{bfseries}) as further load is transferred from the previous failed components. The load that is transferred is also determined by the load-sharing rule. If the system has not failed under this \begin{bfseries}\begin{itshape}Phase I/II cycle\end{itshape}\end{bfseries}, the load is increased and the system eventually undergoes another cycle. This continues until a collection of components fail that cause system failure.

To describe load-sharing rules and the process of system failure, we introduce the notation for a \begin{bfseries}\begin{itshape}breaking pattern\end{itshape}\end{bfseries}. Following Lynch (1999) and Durham and Lynch (2000), the encoding notation for a breaking pattern is a listing in the form
$$p\equiv p_{1}p_{2}\cdots p_{f}\eqno(2.2.1a)$$
where
$$p_{m}\equiv i_{m}(A_{m1}(A_{m2}(\cdots(A_{mk_{m}})\cdots))), m=1, 2, \ldots, f. \eqno(2.2.1b)$$

The breaking pattern in (2.2.1) consists of $f$ failure cycles and $p_{m}$ is referred to as the $m^{th}$ failure cycle. Components $i_{1},i_{2},\ldots , i_{f}$ in the listing are not enclosed in parentheses and indicate components that undergo a Phase I failure. The index $f$ is the number of Phase I/II cycles. Note that an isolated Phase I failure, say $i_{m}$, may initiate no Phase II failures but is still referred to as a Phase I/II cycle, in which case, $p_{m} = i_{m}$.

The parenthetically nested collection of components, the subscripted sets $A$ in (2.2.1b) between the Phase I failures indicate the Phase II failures they initiate. The number of right parentheses indicates the number of Phase II groups of components within a Phase I/II cycle. The groupings indicated by the left parentheses are groupings of components that fail due to load transfer caused by the preceding group's failure.  Note the subscripted sets $B$ and $C$, below, are defined in terms the $A$'s given by the pattern in (2.2.1b).  See Li and Lynch (2011) for details.  (In Hansen et al., 2015, the Phase II failures in a cycle are called a \begin{bfseries}\begin{itshape}burst\end{itshape}\end{bfseries} and the nested parenthetical groupings are referred to as \begin{bfseries}\begin{itshape}inclusive bursts\end{itshape}\end{bfseries}.  An important topic, among many, in this book is the distribution of the size of bursts and that of its behavior (power laws) as a function of the size of bundle.)

Following Harlow et al. (1983), a load-sharing rule is referred to as a \begin{bfseries}\begin{itshape}monotone load-sharing rule\end{itshape}\end{bfseries} if
$$\lambda_{i}(L)\geq \lambda_{i}(M) \mbox{ for all } i\in L, L\subseteq M \subseteq N,$$
and
$$\sum_{i\in M}\lambda_{i}(M)>0, M\subseteq N.$$

Since the load-sharing is monotone, the following system of equations and inequalities for $x$ and $s$ must be satisfied based on the notation from (2.2.1):
$$x_{i_{u}}=\lambda_{i_{u}}(N-C_{u})s_{u}, u=1,\ldots, f \eqno(C1)$$
and
$$\lambda_{i'}(N-C_{u}-B_{u(m-2)})s_{u}<x_{i'}\leq \lambda_{i'}(N-C_{u}-B_{u(m-1)})s_{u} \eqno(C2)$$
for $u=1,\ldots, f$, $i'\in A_{um}$ and $m=1,\ldots, k_{u}$ where $B_{u(-1)}= \emptyset$.

In (C1) write
$$a_{i_{u}}=\lambda_{i_{u}}(N-C_{u}) \eqno(C3a)$$
and let
$$L_{i'}=\lambda_{i'}(N-C_{u}-B_{u(m-2)}) \mbox{ and } U_{i'}=\lambda_{i'}(N-C_{u}-B_{u(m-1)}) \eqno(C3b)$$
Then, from (C3b) the two bounds for the strength of component $i'$ in (C2) can be written as
$$L_{i'}S_{u}<x_{i'}\leq U_{i'}s_{u}. \eqno(C4)$$

Now consider a load-sharing system under increasing load per component, $s$, consisting of $n$ components with independent component strengths $X_{1}, X_{2}, \ldots , X_{n}$. Let $C$ denote the random variable for the number of cycles for component strengths $X_{1}, X_{2}, \ldots , X_{n}$ when the system is a parallel system.  Let $S_{1}<S_{2}< \cdots <S_{C}\equiv S$ denote the sequence of phase I breaking strengths where $S$ is the strength of the parallel system. Let $\{1,\ldots, n\}\equiv P_{0}\supset P_{1}\supset \ldots \supset P_{C-1}$ denote the sets of components that survive the $i^{th}$ cycle, $i = 1,\ldots ,C-1$. Then, since the load-sharing rule is monotone,
$$S_{c+1}=\min\limits_{i\in P_{c}}\frac{X_{i}}{\lambda_{i}(P_{c})} \mbox{ for } c=0, \ldots ,C-1.$$

We first use (C3) and (C4) to derive the joint density of all Phase I breaking stresses, denoted by $f_{\underline{s}}(\underline{S})$, for a system of $n$ iid unit exponential components. We proceed as follows. Pick one of the attainable breaking patterns, say $p$. (Certain patterns may not be attainable.  See Li and Lynch, 2011, for details.)  By (C3a\&b), the joint density of $S$ and $Q$ where $Q$ denotes the random breaking pattern is
\begin{align}
f_{\underline{S}, Q}(\underline{s}, p) &= P(Q=p \mid \underline{S}=\underline{s})dF_{\underline{S}}(\underline{s})\nonumber\\
                &= \prod \limits_{\substack{u=1,\ldots, f\\ m=1, \ldots, k_{u}\\ i' \in A_{um}}}[F_{i'}(U_{i'}s_{u})- F_{i'}(L_{i'}s_{u})][\prod_{u=1}^{f}a_{i_{u}}f_{i_{u}}(a_{i_{u}}s_{u})]d\underline{s} \tag{2.2.2}
\end{align}
where we adopt the convention throughout that terms in the first product in (2.2.2) are equal to 1 when $A_{um} = \emptyset$.

Note that (2.2.2) is similar in form to (2.1.1a).  Thus, if the fiber strengths are iid unit exponentials, we get similar results to (2.1.2-5) from arguments similar to (2.1.1b-e).  For example, the strength distribution of the bundle is a mixture over the scale parameter of gammas with shape parameter $n$.  The mixing distribution itself is a mixture where the components in the mixing distribution are size-biased convolutions of uniforms where the uniforms are based on the load-sharing rule and the failure pattern associated with that component.

For a load-sharing system with arbitrary structure, let $P^{*}$ denote the set of path sets for the structure.  It is assumed without further comment that the systems are coherent; i.e., systems for which all components are relevant in the sense that a component working or not affects whether the system works. Coherent systems can be described in terms of \begin{bfseries}\begin{itshape}path\end{itshape}\end{bfseries} and \begin{bfseries}\begin{itshape}cut\end{itshape}\end{bfseries} sets. Recall (see Barlow and Prochan, 1975) that a path (cut) set for a system is a set of components where, if all the components in a path (cut) set work (fail), the system works (fails). Thus, the system works (fails) if and only if at least one path (cut) set works (fails).

Given $X_{1}, X_{2}, \ldots, X_{n}$, the collection $\{P_{c}\}$ is random. The structure given by $P^{*}$ still works if and only if the set of unfailed components is $P_{c}\in P^{*}$.  Thus, the strength of the structure is $S^{*}=\max\limits_{P_{c}\in P^{*}}S_{c+1}=\max\limits_{P_{c}\in P^{*}}\min\limits_{i\in P_{c}}\frac{X_{i}}{\lambda_{i}(P_{c})}$.  Because $P_{0}\supset P_{1}\supset \ldots \supset P_{C-1}$, $S^{*}=S_{c^{*}+1}$  where  $P_{c^{*}}$ is the smallest path set of the system structure among $P_{0}\supset P_{1}\supset \ldots \supset P_{C-1}$.  It follows that $P_{c}$ is a path set for $c \leq c^{*}$ and the complement of $P_{c^{*}+1}^c$ has to be a cut set of the structure.  Let $A^{c}$  and $|A|$ denote the complement and the cardinality, respectively, of the set $A$.  For fiber strengths that are iid unit exponentials, the strength distribution of the bundle is a mixture over both the scale and shape parameter of gammas where the shape parameter is between $P_{c^{*}}^c$  and $P_{c^{*}+1}^c$. See Section 3 of Durham and Lynch (2000) for a complete discussion regarding the value of the shape parameter.  The mixing distribution again is a mixture where the components in the mixing distribution are size-biased convolutions of uniforms.

\noindent \begin{bfseries}Comments:\end{bfseries}\\
\begin{bfseries}(i)\end{bfseries}  Extensions: Let $X_{1}, X_{2},\ldots, X_{n}$ be independent Weibull strengths where the survival distribution of $X_{i}$ is $\overline{W}(x; \sigma_{i},\rho)=exp\{-(\frac{x}{\sigma_{i}})^{\rho}\}$. Consider a load-sharing system of these components with load-sharing rule $\{\lambda_{i}(M): i \in M, M\subseteq N \}$. Then, since $Z_{i}\equiv (\frac{X_{i}}{\sigma_{i}})^{\rho}$, $i = 1,2,\ldots, n$ are iid unit exponentials, the previous threshold representations for iid unit exponentials can be modified by replacing the load-sharing rule $\{\lambda_{i}(M)\}$ by $\{(\frac{\lambda_{i}(M)}{\sigma_{i}})^{\rho}: M\subseteq N \}$ and a change of variable from $x$ to $x^{1/\rho}$. \\
\begin{bfseries}(ii)\end{bfseries} Interpretation of the Parameters in the Threshold Representations:  For Weibull strengths, the distribution representations’ scale parameters are in terms of the load-sharing rule and contains the mechanics regarding the bundle.  On the other hand, the shape parameters contain the information regarding the number of fiber failures prior to the catastrophic failure of the bundle and is somewhat analogous to crack growth and critical crack size in metals.\\

\section {The Gibbs Distribution of the States of the Fiber Bundle Model}

Below is a summary of results from Sections 2 and 4 in Li et al. (2018).  In this formulation, let $N = \{1,\ldots, n\}$ denote a set of sites/nodes.  These nodes can either be occupied or empty.  The set $A\subseteq N$ of occupied nodes will be referred to as a \begin{bfseries}\begin{itshape}configuration\end{itshape}\end{bfseries} and we are interested in modeling the distribution of the configuration.  In our contexts, the configurations are sets of working components in the fiber bundle.

One family of models are the \begin{bfseries}\begin{itshape}Gibbs measures\end{itshape}\end{bfseries}.  A real valued function $U(A)$, $A\subseteq N$ with $U(\emptyset)=0$ , will be called the \begin{bfseries}\begin{itshape}energy\end{itshape}\end{bfseries} on $N$. The \begin{bfseries}\begin{itshape}Gibbs measure with energy\end{itshape}\end{bfseries} $U$ is given by
$$P(A)=\frac{\exp \{-U(A)\}}{Z} \mbox{ for } A\subseteq N \eqno(3.1a)$$
where $Z$ is a normalizing constant
$$Z=\sum \limits _{A\subseteq N} \exp \{-U(A)\} \eqno(3.1b)$$
referred to as the \begin{bfseries}\begin{itshape}partition function\end{itshape}\end{bfseries} in statistical mechanics and $P(\emptyset)=Z^{-1}$.

If $U$ is an energy function on $N$, define $V(A)$ for $A\subseteq N$ by
$$V(A)=-\sum \limits_{B\subseteq A}(-1)^{|A-B|}U(B) \eqno(3.1c)$$
where $|A|$ denotes the cardinality of the set $A$.  We have immediately
$$U(B)=-\sum \limits_{A\subseteq B}V(A) \eqno(3.1d)$$
since (3c\&d) are just the \textit{\textbf{M\"{o}bius inversion formulas}} that relate $U$ and $V$.  We will call $V$ the \begin{bfseries}\begin{itshape}potential\end{itshape}\end{bfseries} corresponding to energy $U$.

It is implicit in equations (3.1) that the Gibbs measure satisfies the \textit{\textbf{positivity condition}}:
$$P(A)>0 \mbox{ for } A\subseteq N. \eqno(3.2)$$
Since we can define the energy by $U(A)\equiv \log P(A) - \log P(\emptyset)$ for any probability measure satisfying the positivity condition, we see that such a measure is a Gibbs measure.

In this case, the \begin{bfseries}\begin{itshape}local structure\end{itshape}\end{bfseries} for the Gibbs measure on the subsets of $N$ is determined by the log odds ratio of site $i$ being occupied to that it is not given $A \backslash \{i\}$, i.e.,
$$\sigma_{i}(A)\equiv \log \frac{P(A)}{P(A\backslash \{i\})}.\eqno(3.3)$$
Thus, from (3.3), (3.1a\&b) and (3.1d) where in (3.1d),
\begin{align}
 \sigma_{i}(A)&= \log \frac{P(A)}{P(A\backslash \{i\})}=U(A\backslash \{i\})-U(A) \nonumber\\
              &= \sum \limits _{K\subseteq A}V(K)- \sum \limits _{K\subseteq A\backslash \{i\}}V(K)\tag{3.4}\\
                &= \sum \limits _{K\subseteq A: i\in K}V(K). \nonumber
\end{align}

Identity (3.4) gives a way to calculate the log odds ratios by summing potentials. The following theorem shows, via the M\"{o}bius inversion formula, how to obtain the potentials through the log odds ratios.

\noindent \begin{bfseries}Theorem 3.1\end{bfseries} (\begin{itshape}Li et al., 2018, Theorem 2.1\end{itshape}) Let $\sigma(A)\equiv \sum \limits_{i\in A}\sigma_{i}(A)$. Then, for $K\neq \emptyset$,
$$V(K)=\frac{\sum \limits _{A\subseteq K}(-1)^{|K \backslash A|}\sigma(A)}{|K|} \eqno(3.5)$$

The starting point to define the Gibbs measure for load-sharing systems with component strength distributions is (3.3) since the log-odds at a load per component $s$ for the $i^{th}$ component for configuration simply is
$$\sigma_{i}(A,s)=\log \frac{\overline{F}_{i}(\lambda_{i}(A)s)}{F_{i}(\lambda_{i}(A)s)}. \eqno(3.6)$$

Let $\sigma(A,s)\equiv \sum \limits _{i\in A}\sigma_{i}(A,s)$. Then, it follows from (3.5) and (3.1d), how the potentials, $V(K,s)$, and energy,  $U(A,s)$, are determined by these log-odds. In which case, the Gibbs measure for a load-sharing system is $P_{s}(A)=\frac{exp\{-U(A,s)\}}{Z(s)}$, for $A\subseteq N$.

The Gibbs measure, $P_{s}$, is determined by a large number of parameters/functions, $\{V(K,s)\}$. Later, in our analysis of Rosen specimens we will see that the structure of the fitted $P_{s}$ for the $4\times 4$ Rosen grid can be approximated by much simpler Gibbs measures.\\

\section{A Description of Rosen's Experiment:  The Chain-of-Bundles Model}
In this section we discuss Rosen's nine Series A specimens. Each specimen had a test section that was ``$0.5\times 1$ inch in size and 0.06 inch thick" and contained ``90-100 parallel glassfibers of 0.005 inch diameter" (Rosen 1964, p. 1990).  All nine specimens were tested under increasing load, seven until they failed and two until a load at which the testing device malfunctioned.  Specimen A-7 consisted of 93 fibers and is given special status since a sequence of photographs were taken until it failed at an ultimate load (UL) of 116 pounds.  Besides discussing various aspects of the analysis of the specimen data in Sections 4, some technical background material needed regarding partitioned based priors, Weibull plots, chain of bundle asymptotics and absorbing state load-sharing rules for this discussion are also given.

\subsection{The Chain-of-Bundles Model}
Quoting from Li et al. (2018) ``W. B. Rosen (1964, 1965) conducted some remarkable experiments on unidirectional fibrous composites that gave seminal insights into their failure under increasing tensile load. He discovered that no load could be borne through a fiber around a fiber break in a composite for a distance he referred to as the \begin{bfseries}\begin{itshape}ineffective length\end{itshape}\end{bfseries}. This discovery led to a grid system of components conceptualization where the nodes in the grid are ineffective length fiber components. The component strengths are independent and a component/node fails when the load at the node exceeds the component strength. The tension/interaction between the loads and the component strengths creates dependencies between the nodes."

``This conceptualization also led Rosen to model the composite as something he called a \begin{bfseries}\begin{itshape}chain-of-bundles model\end{itshape}\end{bfseries}. The \begin{bfseries}\begin{itshape}chain\end{itshape}\end{bfseries} is a series system of parallel subsystems of nodes which he referred to as \begin{bfseries}\begin{itshape}bundles\end{itshape}\end{bfseries}. The bundles were horizontal collections of nodes across the grid and the chain fails when one of the bundles fails and where a bundle fails when all the components/nodes in the bundle fail. No load is transferred between bundles but there is load-sharing between nodes in a bundle described by a load-sharing rule. Originally he used the equal load-sharing rule where load is carried equally within a bundle and load from a failed component in a bundle is transferred equally to unfailed nodes/components within a bundle. But, he and later authors proposed more localized rules, some based on crack growth considerations, but only in horizontal bundles."

\subsection{The Analysis of Specimen A-7}
``Grego et al. analyzed the photographs of specimen A-7 from one of Rosen's experiments. (Copies of these photographs are given there). For their analysis, they used a $22\times93$ grid system of components to estimate the component strength distribution since there were 93 fibers in the specimen where each fiber was divided into 22 fiber segment components based on the median ineffective length of the specimen."  In Figures 4 and 5, there, failures in the grid system correspond to those in the photographs.

``In the experiments, photographs were taken to identify fiber fractures at a series of percentages of the ultimate load. In the initial experiment, polarized transmitted light was used. At zero load, the fibers are dark and the binder between fibers appears light. As the load increases, the fibers appear brighter indicating increased load, and breaks appear as dark rectangular regions at random locations. (The dark rectangular regions are determined by the ineffective length and their being dark indicates that no load is borne in the region.)"

``The ``X" shaped area of increased brightness around a break suggests that load is transferred through the matrix material to unbroken fiber components that are diagonally and horizontally (but not vertically) adjacent to the break. This means that a local load-sharing rule would be appropriate to model load transfer where the load at a break is transferred equally in the six diagonal and horizontal directions as depicted in Figure 2. See Grego et al. (2014) for further details."

The above description of the load-sharing sufficed for Grego et al.'s analysis but does not well-define the load-sharing rule at this point and was not needed for their analysis. This is needed later, though, to fit the Gibbs measure for the bundles in the chain-of-bundles model for Rosen's specimen data and discussed in the next subsection about absorbing state load-sharing rules.

The analysis in Grego et al. (2014) was based on a nonparametric Bayes procedure where the prior was a Dirichlet process prior.  The Dirichlet process parameter is a finite measure $\alpha(\cdot)$ on $[0, \infty)$. The normed Weibull distribution with shape parameter 6.5 and scale parameter 1.5 and weight/norming constant $\alpha([0, \infty))=50$ was chosen for $\alpha(\cdot)$.  The choice of these values for the shape and scale parameters were based on some apriori information cited in the paper.

In an ideal situation, the Phase I breaking strengths could be measured but the Phase II breaking strength are always censored and only known to be in a set of values.  Since the data for Specimen A-7 is a series of photographs where component failures are only known to occur in a given photograph, all breaking strengths are censored.  In such censoring situations, Dirichlets priors can be converted to partitioned based Dirichlets (PBD) where the partition is determined by the censoring sets (Sethuraman and Hollander, 2009).  The data from Rosen's photographs results in a partition consisting of 22 intervals (not to be confused with the 22 rows in the grid for the composite) based on the censoring.

In this situation, because all the observations are censored, the partition probabilities are just multinomial probabilities where the Dirichlet process over a partition set is just a smoothing method for the posterior probability of the strength distribution estimator over that set.  When there is no censoring in a multinomial problem with $k$ cells (equivalent to a partition of size $k$) with a Dirichlet prior, the posterior is Dirichlet where the posterior parameter is the prior parameter updated by one for the cell that the observation appears in.  When an observation is censored, e.g., say it is only known to occur in the first two cells, the prior is used to assign the observation to cell 1 or cell 2.  If it assigns it to cell 1 the prior parameter for cell 1 is updated by one while the assignment to 2 updates the prior parameter for cell 2.  It follows that the posterior is then just a mixture of two Dirichlets with these two updated parameters where the mixing proportions are just the assignment probabilities from the prior Dirichlet.  In general, the natural (conjugate) prior for these types of nonparametric Bayes problems is a PBD where the posterior is just a PBD.  The formal derivation of these results can be found in Sethuraman and Hollander, (2009).

The main finding of the analysis in Grego et al. (see Figure 9 there) was that a Weibull with $\rho=5$ and $\sigma=2$ was a reasonable model for the lower tail of the component strength distribution of an ineffective length fiber.  In Section 4.5, we discuss how Li et al. (2018) use this information to see that a $4\times4$ grid is an appropriate size bundle to model Rosen's specimen data as a chain-of-bundles model and to determine the Gibbs measure for the of this size. We also discuss their nonparametric estimation of the specimen breaking strength distribution for Rosen's 9 specimen breaking strengths.  This estimate and its confidence bands are used to judge the validity of the chain-of-bundles approximations given later.

The next two subsections give some necessary background material needed for this discussion.  In Section 4.3, absorbing state load-sharing rules are used to well-define the local load-sharing rule used in Grego et al.'s analysis and, in Section 4.6, to derive the Gibbs measure and investigate its behavior.  Weibull plots and chain-of-bundle asymptotics used in Section 4.5 are given in Section 4.4.

\subsection{Absorbing State Load-Sharing Rules}
Consider a network of components.  A network is defined by a directed graph $\textbf{\textit{G = G(N,E)}}$ where $\textbf{\textit{N}} ={1,2,\ldots,n}$ and $\textbf{\textit{E}}$ are the sets of nodes/components and edges, respectively, in $\textbf{\textit{G}}$.  In essence, the graph is the basis of a transition diagram for the Markov chain which will be used to calculate the load-sharing rule.  Let $\underline{\textbf{\textit{P}}} =\{p_{i,j}: i \mbox{ and } j =1,\ldots,n\}=\{p_{i,j}\}$ be a one-step transition probability matrix on the nodes $\textbf{\textit{N}}$ of $\textbf{\textit{G}}$ where there is a directed edge from $i$ to $j$ in $\textbf{\textit{G}}$ if $p_{i,j} > 0$.

The state of a node is either 1 or 0 indicating that the node works or has failed.  Each component $i$ has a strength $S_i$. Let $s > 0$ denote the load per $n$ components that is imposed on the network and let $A$ denote the set of working components. Then, as before, the load at component $i\in A$ is given by $\lambda_{i}(A)s$  and the component fails if $\lambda_{i}(A)s<S_i$.  The load is determined by a load-sharing rule referred to as \textbf{\textit{absorbing state load-sharing rules}} defined in terms of an absorbing state Markov chain on the graph $\textbf{\textit{G}}$.  The notion of an absorbing state load-sharing rule is now formalized and shown to be monotone.

For a set of working components $A$ let $^{A}\underline{P} =\{^{A}p_{i,j}\}$ where, if $i \in A$, $^{A}p_{i,j} = 1$ when $j=i$ and $^{A}p_{i,j} = 0$ if not, and $^{A}p_{i,j} = p_{i,j}$ if $i$ and $j$ in $A^{c}$.  By rearranging the rows and columns of $^{A}\underline{P}$ it can be rewritten as the partitioned matrix

\[{}^A\underline{P} = \left\{ \begin{array}{ccc}
{}^A\underline{I} & {}^A\underline{0} \\
{}^A\underline{R} & {}^A\underline{Q} \end{array} \right\}\]

\noindent where $^A\underline{I}$ is the $|A|\times|A|$ identity matrix, $^A\underline{Q}$ is the subprobability matrix of one-step transition probabilities from states in $A^c$, and $^A\underline{R}$ is the matrix of one-step transitions from states in $A^c$ to those in $A$. Then, the absorption probabilities $\{^AU_{i,j}:i\in A^c, j\in A\}$ satisfy the system of equations
$$^Au_{i,j}=\sum\limits_{k\in A^c}{}^Aq_{i,k}{}^Au_{k,j}+{}^Ar_{i,j} \eqno(4.3.1a)$$
that can be written in matrix form as
$$^A\underline{U}= {}^A\underline{Q}{}^A\underline{U}+{}^A\underline{R}.\eqno(4.3.1b)$$
Identity (4.3.1) has solution
$$^A\underline{U}= \sum\limits_{m=0}^{\infty}{}^A\underline{Q^m}{}^A\underline{R}\equiv({}^A\underline{I}-{}^A\underline{Q})^{-1}{}^A\underline{R}.\eqno(4.3.2)$$

Interpret $u_{i,j}$ as the proportion of the load at $i\in A^c$ that is transferred to $j\in A$. Then we define an \textit{\textbf{absorbing state load-sharing rule}} by
$$\lambda_{i}(A)=1+\sum\limits_{i\in A^c}{}^Au_{i,j}.\eqno(4.3.3)$$

\noindent \begin{bfseries}Lemma 4.3.1\end{bfseries} An absorbing state load-sharing rule is monotone.

\noindent \textbf{Proof}: Let $A\subset B$, $j\in A$ and $i\notin B$. Since any absorption path from $i$ to $j$ that contributes to ${}^Bu_{i,j}$ is also one that contributes to ${}^Au_{i,j}$, ${}^Bu_{i,j}\leq {}^Au_{i,j}.$ Thus, from (4.3.3),
$$\lambda_{j}(B)=1+\sum\limits_{i\in B^c} {}^B u_{i,j} \leq 1+\sum\limits_{i\in B^c} {}^A u_{i,j}\leq 1+\sum\limits_{i\in A^c} {}^A u_{i,j}
=\lambda_{j}(A). \hspace{3em} \blacksquare$$

\noindent \textbf{Remark:} The transition matrix based on the local load-sharing rule used in the $22\times93$ grid to analyze Specimen A-7 or any of the grids discussed later is defined as follows. Nodes in a grid network are neighbors if they are adjacent diagonally or horizontally in the grid but not vertically. Transition probabilities from any given neighbor to an adjacent neighbor is 1 over the number of neighbors. This gives the one step transition matrix $\underline{P}$ that is used to define the load-sharing rule given by (4.3.3).

\subsection{Chain-of-Bundles: Weibull Plots and Minimums}
Let $\overline{F}$ be a survival function. A Weibull probability plot of $\overline{F}$ is the graph $\{(\ln x, \ln [-\ln \overline{F}]):x\geq 0 \}$. The Weibull probability plot of the Weibull survival function $\overline{W}(x,\sigma,\rho)=\exp \{-(x/\sigma)^\rho\}$ is the linear graph $\{(\ln x, \rho \ln x - \rho \ln \sigma):x\geq 0\}$ with slope, the shape parameter
$\rho$, and intercept $-\rho  \ln \sigma$ at $x=1$.

We now consider the behavior of the slope in a Weibull plot as $x\rightarrow 0$. Let $F(x)\sim Kx^\beta$. Since $-\ln \overline{F}(x)=F(x)+o(F(x)) \sim Kx^\beta$, $\ln[-\ln \overline{F}(x)]-\beta \ln x - \ln K \rightarrow 0$ as $x \rightarrow 0$. So, the slope of the Weibull plot behaves like $\beta$ as $x\rightarrow 0$. This has implications for the minimum from a distribution $F(x) \sim Kx^\beta$. The following lemma shows that the asymptotic distribution of the minimum is Weibull with shape parameter $\beta$.

\noindent \begin{bfseries}Lemma 4.4.1\end{bfseries} Let $M_r$ be the minimum of $r$ observations from $F(x)\sim Kx^\beta$. Then,
$$P(r^{1/\beta}M_r >x)\xrightarrow{r\rightarrow \infty} \exp \{-Kx^\beta\}.$$

\noindent \textbf{Proof}: From an argument similar to the one above in the preceding paragraph,
$$\ln P(r^{1/\beta}M_r >x)=r\ln \overline{F}(x/r^{1/\beta})\sim -Kx^\beta. \hspace{3em} \blacksquare$$

From Section 2.2, the strength survival distribution for a parallel load-sharing system of $n$ iid unit exponential strengths is, as $x\rightarrow 0$,
$$F(x)=\sum\limits_{j=n}^\infty \int \frac{(\theta x)^j e^{-\theta x}}{j!}a(\theta)d(\theta)=\frac{x^n}{n!}\int \theta ^n a(\theta)d\theta +o(x^n)\equiv Kx^n + o(x^n)\eqno(4.4.1)$$
where the first identity follows from the threshold representation and the relationship of the gamma distribution with integer shape parameter and the Poisson distribution. Thus, from Lemma 4.4.1, for a chain of bundles of $n$ parallel iid unit exponential components, the chain strength distribution is asymptotically Weibull with scale parameter $K=\int \frac{\theta^n}{n!} a(\theta)d\theta$  and shape parameter $n$.

For $n$ independent Weibull strengths with the same shape parameter, $\rho$, but with different scale parameters, $\sigma_i$, $i=1,2,\ldots,n$, we get the following from Comment (ii) in Section 2.2 by replacing the load-sharing rule $\{\lambda_i(M)\}$ by $\{(\frac{\lambda_i(M)}{\sigma_i})^\rho : M\subseteq N \}$ and changing $x$ to $x^\rho$ in (4.4.1):			
$$F_W(x)=\sum\limits_{j=n}^\infty \int \frac{(\theta x^\rho)^j e^{-\theta x^\rho}}{j!}a_W(\theta)d(\theta)=\frac{x^{\rho n}}{n!}\int \theta ^n a_W(\theta)d\theta +o(x^{\rho n})\equiv K_W x^{\rho n} + o(x^{\rho n}).\eqno(4.4.2)$$
Here $a_W(\theta)$ is the mixing distribution based on the load-sharing rule $\{(\frac{\lambda_i(M)}{\sigma_i})^\rho : M\subseteq N \}$.

For a bundle with arbitrary structure and fiber strengths that are iid unit exponentials, from the discussion in Section 2.2, the strength distribution of the bundle is a mixture over both the scale and shape parameter of gammas where the shape parameter is between $|P_{c^*}^c|$ and  $|P_{c^*+1}^c|$.  It is easy to see that (4.4.1) holds where $n$ is replaced by the minimum shape parameter, say $m$, and the constant $K$ is based on the conditional $m^{th}$ moment of the mixing distribution of the scale parameter given the shape parameter is $m$.  Similarly, for independent Weibull strengths with the same shape parameter, $\rho$, but with different scale parameters, (4.4.2) holds with $n$ replaced by $m$. Here the support of the shape parameter in the mixing distribution is contained in $\{\rho,2\rho,\dots,n\rho \}$.

Thus, combining (4.4.1) and (4.4.2) and the above comments with Lemma 4.4.1, we get the following regarding the asymptotic distribution for the chain-of-bundles model.\\

\noindent \begin{bfseries}Lemma 4.4.2\end{bfseries} Consider a load-sharing bundle of $n$ independent Weibull strength components with the same shape parameter, $\rho$, with a given system structure. Let $k$ denote the size of the smallest cut set(s) for the structure. Then, the asymptotic distribution of the strength of the chain is Weibull with shape parameter  $k\rho$ and the scale parameter based on the $k^{th}$ moment of the mixing distribution of the scale parameter given that the shape parameter is $k\rho$. \\

Thus, for large chains, the \textbf{inflation factor}, $k$, the asymptotic chain Weibull shape parameter is the critical number of failures in the bundle that cause catastrophic failure of the chain.

\subsection{Modeling the Specimen Breaking Strength Distribution as a Chain-of-Bundles}
In this section, we summarize Li et al.'s (2018) investigation of the chain-of-bundles model with grids for bundles as a model for the breaking strengths of Rosen's Series A specimens. The data for these specimens is in Table 2 there.  Since there are two right censored observations, the Kaplan-Meier (KM) estimator was used to construct the cdf plots and the Weibull plots in Figure 11 there. The $95\%$ conﬁdence bands in the plots are based on the KM estimator and the solid reference lines are for the Weibull with shape and scale parameters $\rho_s = 22.04$ and $\sigma_s = 117.69$, respectively, which are the values of the maximum likelihood estimates (mle) of these parameters. Here, the subscript, $s$, is for specimen.

Simulations of size $10,000,000$ were used to determine the grid strength distribution for various size grids. The component strength distribution was Weibull with $\rho = 5$ and $\sigma = 2$ based on the discussion in the Section 4.2.2.

A reason for studying such grids is that they accommodate local load-sharing in not just the horizontal direction. Zweben and Rosen (1970) and Zhao and Takeda (2000b) discuss local load-sharing rules to define the size of small horizontal clusters (see Figure 1 in the latter) of ineffective length fibers whose failure causes catastrophic failure of the composite. Zhao and Takeda found that the Zweben-Rosen local load-transfer model would result in critical horizontal clusters of size $2-4$ but the size would be larger for more realistic composites that are based on the Curtin-Takeda load-transfer model. Again, Zhao-Takeda and Zweben-Rosen were both studying horizontal clusters while the grids studied here accommodate the load-transfer seen in Rosen's experiments which is both in the horizontal and diagonal direction. Note, though, that if the Zweben-Rosen model for three dimensional unidirectional fibrous composites, where the load-transfer around a break is to the four adjacent fiber components, is modified to a planar unidirectional composite with four adjacent components (say, just diagonally), then their load-transfer model is close, but more severe, to the model used here where the adjacency is 4 diagonal and 2 horizontal components. It is not surprising that the Zweben-Rosen model would slightly underestimate the critical number of failures, 4, in the analysis given here.

The distributions and their Weibull plots are given in Figures 12-14 in Li et al., (2018), for $3\times 3$, $4\times 3$, $5\times 3$, $6\times 3$, $4\times 4$, $5\times 5$, $6\times 6$, $3\times 4$, $3\times 5$, and $3\times 6$ grids where ``The slope of the lower tails ($s \leq 1$) in the Weibull plots (Figures 12(b), 13(b) and 14(b)) indicates that a chain of grid bundles for these size grids will be asymptotically Weibull where the slope, $\rho_g$, is the Weibull shape parameter for the given grid. The relative slope, $k = \rho_g/\rho$, is the inflation factor (IF) where the extreme value Weibull ``asymptotics" discussed in the introduction of Section 4.2" (see the previous subsection here, especially Lemma 4.4.2) `` suggest that $k$ is the critical number of components that cause catastrophic failure of the grid for $s \leq 1.$".

In Figure 1 below we give graphs of the distributions and their Weibull Plots and of the Weibull fit for the specimen data and of the chain-of-grid models for the square grids, $c\times c$, $c = 4,5,6$ and for the $r\times 3$ grids, $r=3,4,5,6$. You can see a distinct difference between the square grids graphs and plots which are indistinguishable from the mle Weibull fit and the latter rectangular grids which ``underestimate" the mle fit since they are, for the most part, to the left of it. Figure 17 of Li et al. (2018) shows that, even though the breaking strength distributions of chains of the $3\times4$, $3\times5$ and $3\times6$ grids fall within the confidence bands, they considerably overestimate the strength for the mle Weibull fit.

\begin{figure}[ht]
\centering
\subfigure[Distributions]{
\includegraphics[width = 3.0 in ]{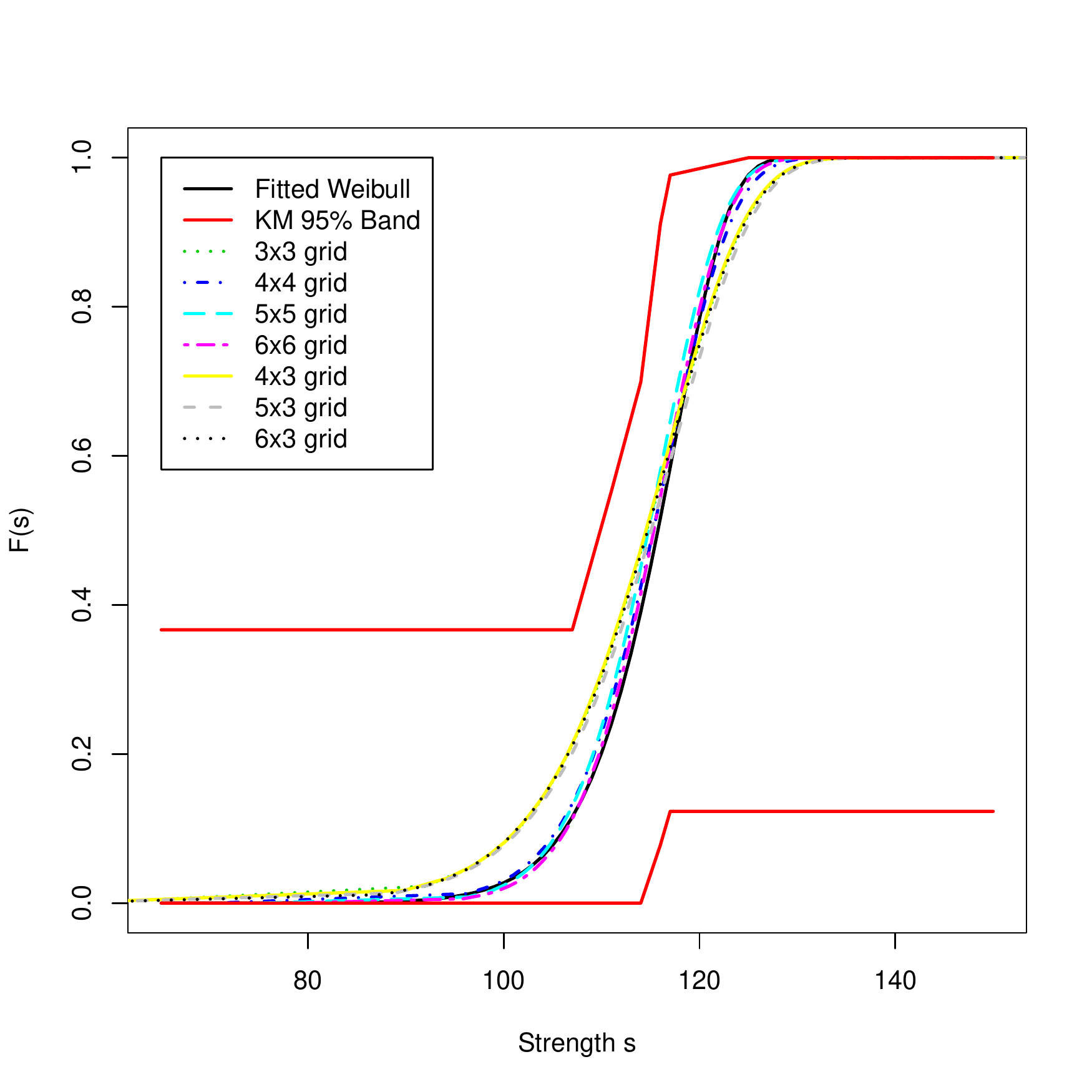}
}
\subfigure[Weibull Plots and Weibull Fit]{
\includegraphics[width = 3.0 in ]{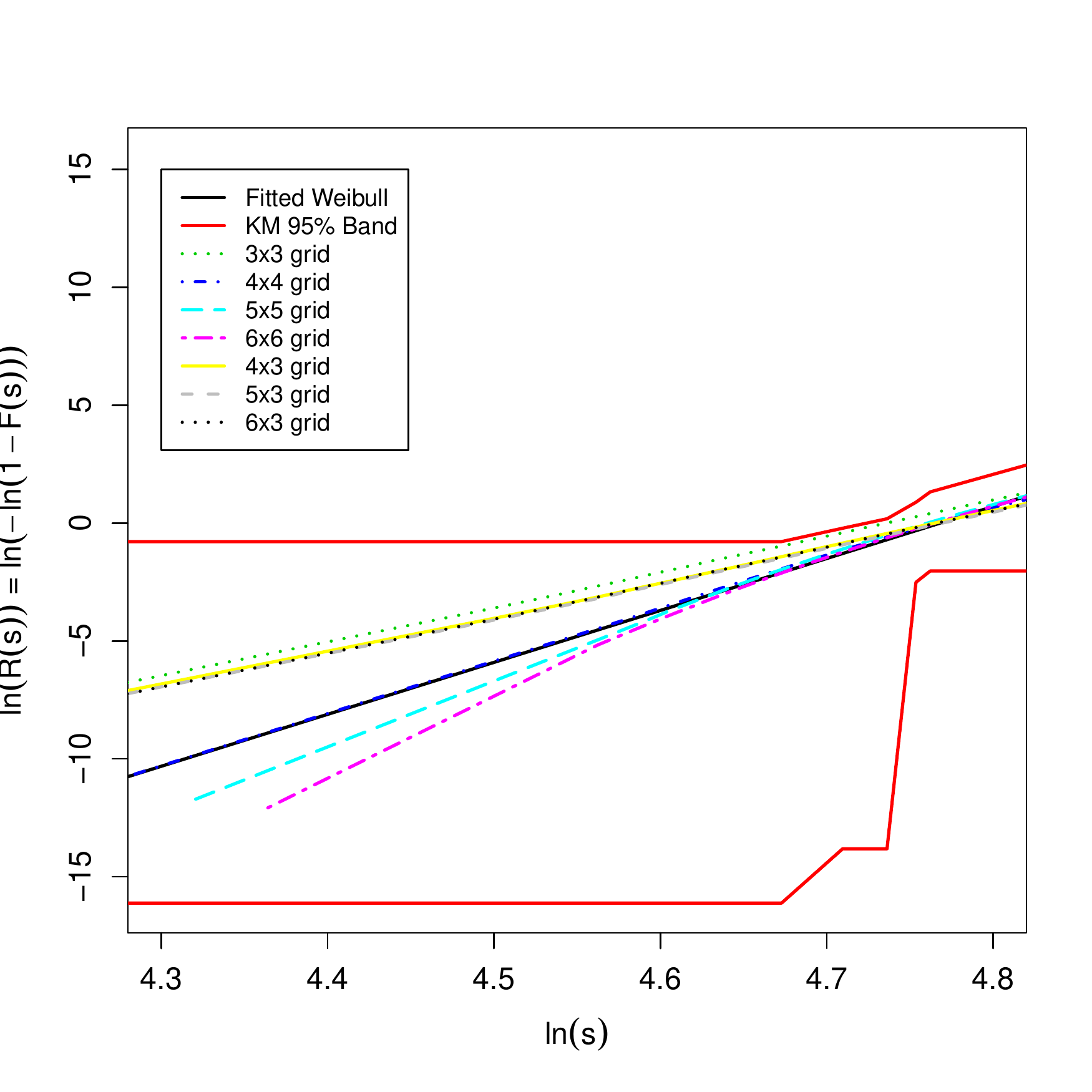}
}
\caption{Distributions, Weibull Plots, and the Weibull Fit for the Specimen Data and of the Chain of Grids.}
\end{figure}

\subsection{The Gibbs Measure for the $4\times 4$ Grid}
We now give a condensed version of the findings in Section 4.2.1 of Li et al. (2018) regarding the Gibbs measure for the $4\times 4$ grid. Here we make use of the material in Section 3 about Gibbs measures.  Define the load-sharing rule from the remark at the end of Section 4.3 and (4.3.3). From this, we can define the log-odds $\sigma_i(A,s)=\log \frac{\overline{F}_i(\lambda_i(A))}{F_i(\lambda_i(A))}$ and $\sigma(A,s)=\sum\limits_{i \in A}\sigma_i(A,s)$ at a given load per component $s$.  Using this with (3.5) and (3.1d) we can define the potentials and energy that defines the Gibbs measure in (3.1a).

Although there are a large number of parameters/potentials, they are well behaved.  This is indicated in Figures 4-6, there, and in Figure 2, below, which is a subset of Figure 6 and indicates the linear relationship between the potentials graphed there for different strength percentiles.  Let $s_p$ denote the $p^{th}$ percentile of the bundle strength distribution. The corresponding slope, $a(s_{p'})$, and y-intercept, $b(s_{p'})$, are graphed in Figure 3, below.

\begin{figure}[h!]
\centering
\subfigure[$p$=.001, $p'$=.01, 1, 10, and 30]{
\includegraphics[width = 3.0 in ]{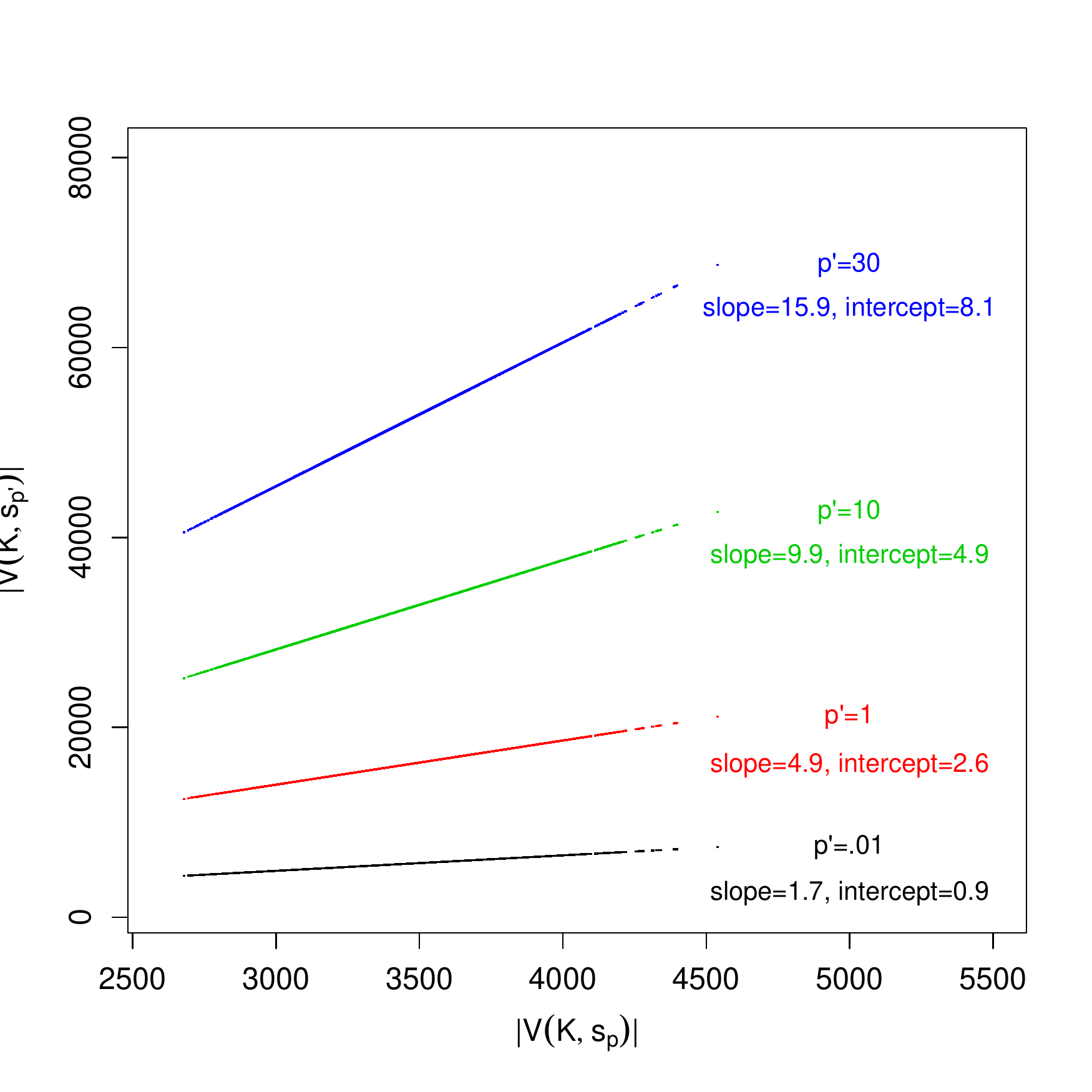}
}
\subfigure[$p$=.001, $p'$= 50, 70, and 90]{
\includegraphics[width = 3.0 in ]{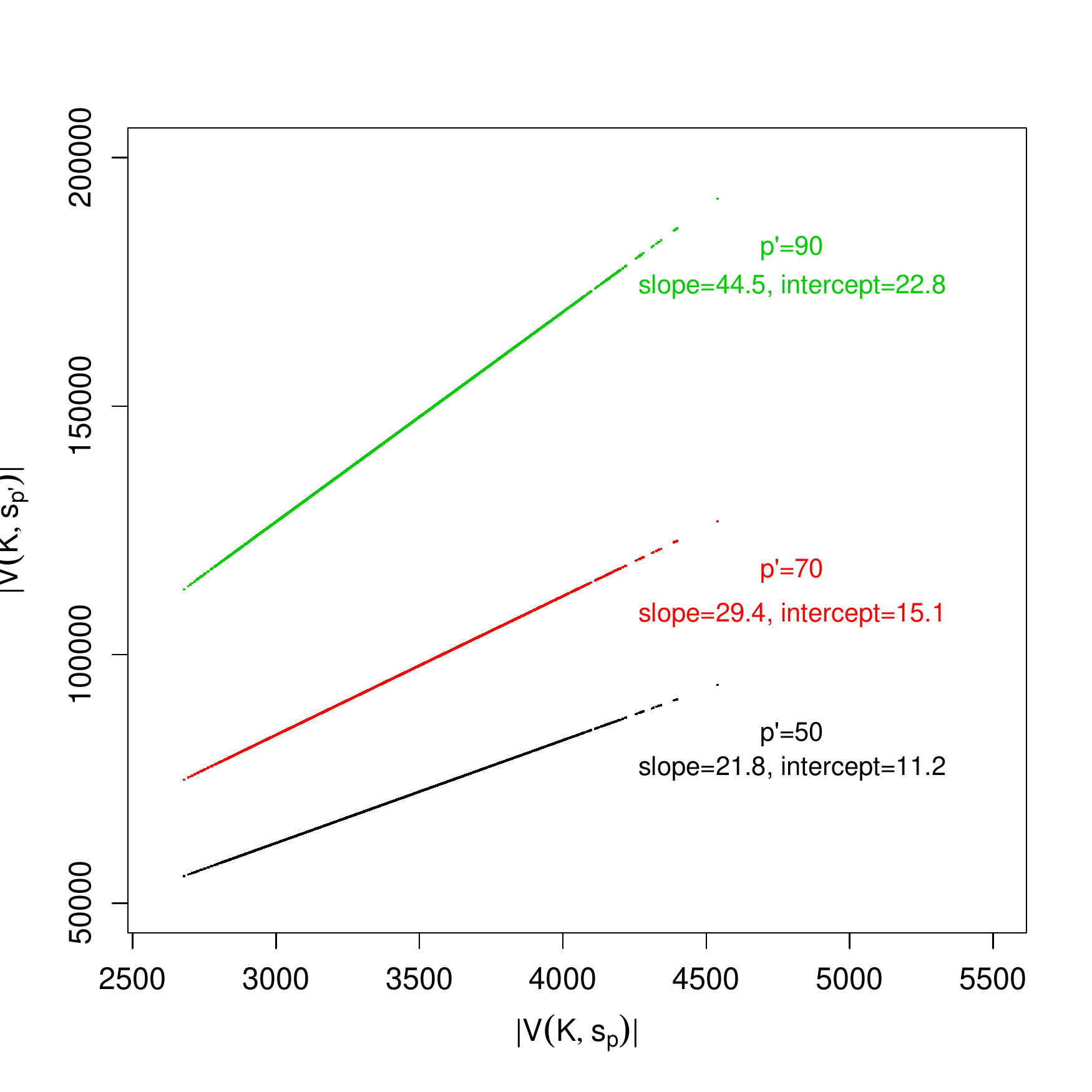}
}
\caption{Potential Scatterplots for System Strength Percentiles $p$=.001 and $p'$=.01, 1, 10, 30, 50, 70, and 90.}
\end{figure}

\begin{figure} [h!]
\begin{center}
\includegraphics [width = 4.0 in ]{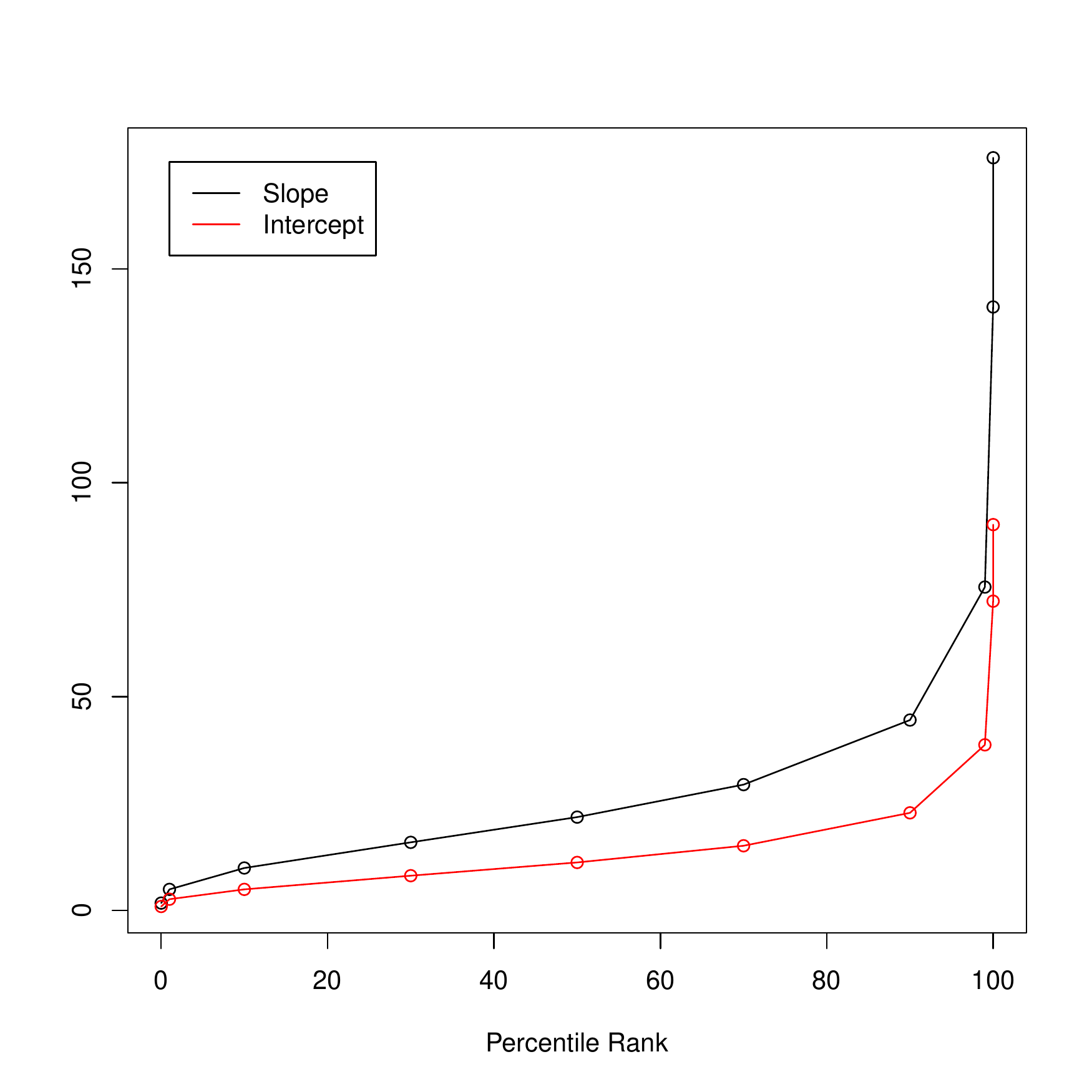}
\caption {Slope and Intercept of the Linear Relationship between the Potentials for System Strength Percentiles $p$=.001 and $p'$=.01, 1, 10, 30, 50, 70, 90, 99, 99.99 and 99.999.}
\label{}
\end{center}
\end{figure}

These figures suggest that there can be a substantial reduction in the parameterization of the Gibbs measure for modeling Rosen's specimen data.  In particular, let $v(k, s_p)$ denote the median potential for sets $K$ for which $k = |K|$.  From Figure 3 and (3.1d), the energy function for the ${p'}^{th}$ is approximately

\begin{align}
 -U(A,s_{p'})&\cong a(s_{p'})\sum\limits_{K\subseteq A}V(K,s_p)+b(s_{p'})2^{|A|-1} \nonumber\\
             &\cong a(s_{p'})\sum\limits_{K\leq |A|} {|A| \choose k} v(K,s_p)+b(s_{p'})2^{|A|-1} \nonumber\\
             &\equiv -a(s_{p'})U_1(A) - b(s_{p'})U_2(A) \nonumber\\
             &\equiv -U_{LMF}(A,s_{p'}) \nonumber
\end{align}
where the second approximation is based on the median approximations of the potentials justified by Figure 5 in Li et al. (2018).

The latter approximation is referred to as the \textit{\textbf{linear median field (LMF)}} approximation by them.  The LMF approximation is a bi-proportional approximation using the two energy functions $U_1$ and $U_2$. The first energy function is totally determined by the load-sharing rule and the component strength distributions while the second is totally determined by the cardinality of the configuration $A$.  Gleaton et al. (2019) have studied implications of Gibbs measures with proportional energies for dry fiber bundles.  How well such implications carry over to this bi-proportional case is subject to further study.

\section{Some Concluding Comments}
Below are two comments regarding the above results.  The first indicates that, for monotone load-sharing systems, the failure process is Markovian.  The second suggests how the previous ideas can be extended to \textit{\textbf{cycles to failure}} (not to be confused with Phase I/II cycles) of a fiber bundle.

(i) Markovian Behavior Under Increasing Load:  The stochastic behavior of a fiber bundle is simple to describe in terms of a repetitive two step Phase I/II Markovian structure (Li, 2009) described as follows.

The first step is a Phase I failure.  The Phase I failure strength is based on the conditional strength distributions of the surviving components after the previous Phase II failures that failed at a load per component, say $s$.  Since the load-sharing rule is monotone, all the relevant information that is needed to determine the Phase I failure in the next phase depends on the set, say $A$, of components surviving after the previous Phase II failures, as well as the conditional strength distributions of the surviving components conditioning on $\lambda_i(A)s$ for the $i^{th}$ component.

Say the next stage Phase I failure is at $s' > s$.  Then, for the next Phase II cycle, the conditional log odds of the $i^{th}$ component surviving load $\lambda_i(A)s'$ given that it survived load $\lambda_i(A)s$ is the same as the unconditional log odds $\sigma_i(A,s')$ given in (3.6) since $\lambda_i(A)s' \geq \lambda_i(A)s$. Thus, the Gibbs measure for the Phase II failures only depends on the Phase I failure at $s'$ and the set of surviving just prior to at $s’$ and no other information.  Note that this formulation has a drawback; it does not keep track of the failure pattern.  The Gibbs measure only determines the set of surviving components in the Phase II cycle.

In terms of \textit{\textbf{Markov random fields (MRF's)}} the local structure of the field is that of the Gibbs measure.  Thus, the set of failed components corresponds to the set of occupied nodes in the network for the MRF.  The graph for the network to analyze Rosen's specimen data is complete so it is not that interesting from a MRF perspective; all the nodes are neighbors.  See Li et al. (2017), Section 4.1, for a discussion of the relationship of the neighborhood structure of the MRF and that of the load-sharing network and for MRF's references.  (This is an earlier version of Li et al., 2018, but with this material on MRF's.)

(ii) Cycles To Failures - Incorporating Damage/Degradation: The following is inspired by Kun et al. (2006) Section 3 formulation of degradation, especially formula (5).  Here consider cycle testing of a bundle under repeated increased load from 0 to $s^*$ in a cycle.  If no damage or degradation is incorporated into the model, then either the bundle fails in the first cycle or it never fails.  This is unrealistic.

To avoid this, one way to account for wear/degradation in cycle testing is to have the strength distributions of the components decrease as the number of cycles increases.  A tractable way to do this is to define the strength distribution of the $i^{th}$ component after the $k^{th}$ cycle to be $F_{ik}(s)=F_i(s/a^k)$ where $0 < a < 1$ is the parameter that accounts for the degradation in strength in a cycle.  We conjecture that much of the above work would apply but needs to be pieced together over the intervals $(0,s^*]$, $(s^*,s^*/a]$, $(s^*/a,s^*/a^2]$, ... where the first, third, ... (the odd numbered intervals) can be viewed as increasing the load over those intervals and where the even numbered intervals have Phase II failures given by degradation from ``jumping" the load from $s^*$ to $s^*/a$ in the second interval etc.  These Phase II failures in the even intervals are not caused entirely by load transfer but are systematic failures due to this ``jumping of load" because of degradation/damage over that cycle.

\section{Summary}
In this paper we have sketched some modeling results and methodology used in the study of fiber bundles and chains-of-bundles.  The modeling results include a threshold representation for the bundle strength distribution, the Gibbs measure for the joint distribution of the states of the bundle and minimum extreme value asymptotics to analyze chain-of-bundles.  The methodology includes partition based priors and Kaplan-Meier estimation to analyze censored data and Weibull analysis methods. These results and analysis methods are illustrated using Rosen's Series A Specimen data.

\end{document}